\documentclass[twocolumn,amsmath,amssymb,aps,prb,floatfix,nofootinbib,superscriptaddress]{revtex4}
\usepackage{graphicx,graphics,times,color}
\usepackage{bm}
\usepackage{longtable}

\def\be{\begin{equation}}
\def\ee{\end{equation}}
\def\ba{\begin{eqnarray}}
\def\ea{\end{eqnarray}}

\def\ra{\rangle}

\begin{document}
\title{Negativity as the Entanglement Measure to Probe the Kondo Regime in the Spin-Chain Kondo Model}
\author{Abolfazl Bayat}
\affiliation{Department of Physics and Astronomy, University
College London, Gower St., London WC1E 6BT, UK}
\author{Pasquale Sodano}
\affiliation{Dipartimento di Fisica e Sezione I.N.F.N.,
Universita' di Perugia, Via A. Pascoli, Perugia, 06123, Italy}
\author{Sougato Bose}
\affiliation{Department of Physics and Astronomy, University
College London, Gower St., London WC1E 6BT, UK}

\begin{abstract}
We study the entanglement of an impurity at one end of a spin
chain with a block of spins using negativity as a true measure of
entanglement to characterize the unique features of the gapless
Kondo regime in the spin chain Kondo model. For this spin chain in
the Kondo regime we determine- with a true entanglement measure-
the spatial extent of the Kondo screening cloud, we propose an
ansatz for its ground state and demonstrate that the impurity spin
is indeed maximally entangled with the cloud. To better evidence
the peculiarities of the Kondo regime, we carry a parallel
analysis of the entanglement properties of the Kondo spin chain
model in the gapped dimerised regime. Our study shows how a
genuine entanglement measure stemming from quantum information
theory can fully characterize also non perturbative regimes
accessible to certain condensed matter systems.
\end{abstract}

\date{\today}
 \maketitle

\section{Introduction}\label{introduction}
 The investigation of entanglement or the truly ``quantum" correlations inherent in many-body condensed matter systems is currently a topic of intense
 activity
 \cite{Amico-RMP,bose-vedral,fazio-osborne,vidal,korepin,Calabrese,Laflorencie-Affleck,Sorensen-Affleck,Lehur}. This
emerging area aims at characterizing many-body states using tools
and measures developed in quantum information. Till date, most
investigations have focussed on either the entanglement between
individual elements, such as single spins, or the entanglement
between two complementary blocks of a many-body system. The former
entanglement is generically non-zero only between nearest or next
to nearest neighbors \cite{bose-vedral,fazio-osborne}. For
complementary blocks, the whole system is in a {\em pure state}
and the von Neumann entropy is a {\em permissible} measure of the
entanglement. In this context much interest has been evoked by
conventional gapless phases, where, due to the absence of an
intrinsic length scale, the von Neumann entropy diverges with the
size of the blocks \cite{vidal,korepin,Calabrese}. In this
backdrop, it is timely to investigate the entanglement in gapless
regimes of a many body system for which its true form and amount
may not be characterized through the entanglement between
individual spins or complementary blocks. Kondo systems are ideal
candidate for these investigations \cite{kondo,Sorensen-Affleck}.

Kondo systems \cite{kondo,Sorensen-Affleck} are expected to be
very distinctive in the context of entanglement for at least two
reasons:  (a) Despite being ``gapless", they support the emergence
of a length scale $\xi$- the so called Kondo screening length
\cite{kondo,Sorensen-Affleck}- which should be reflected in the
entanglement, making it markedly different from that in the more
conventional gapless models studied so far; (b) They are expected
to have a more exotic {\em form} of entanglement than the widely
studied spin-spin and complementary block entanglements. Indeed,
in Kondo systems, the impurity spin is expected to be mostly
entangled with only a specific block of the whole system. This is,
of course, merely an intuition which needs to be quantitatively
verified with a genuine measure of entanglement: this is indeed
the task accomplished in this paper, where we provide the only
characterization of the Kondo regime based entirely on a true
measure of entanglement.

 The simplest Kondo model \cite{kondo, hewson} describes a single impurity spin interacting with the conduction
electrons in a metal; the ground state is a highly nontrivial many
body state in which the impurity spin is screened by conduction
electrons in a large orbital of size $\xi$. Many physical
observables vary on the characteristic length scale $\xi$, which
is a well defined function of the Kondo coupling \cite{kondo}.
Determining the spatial extent of the Kondo cloud has been so far
a challenging problem repeatedly addressed by various means
\cite{Sorensen-Affleck,Frahm,dots}. This includes an investigation which introduces
a quantity called ``impurity entanglement entropy" which, however, is not a measure of entanglement and cannot
{\em quantify} the entanglement within the system \cite{Laflorencie-Affleck,Sorensen-Affleck}.
Recently \cite{kondo-Affleck},
it has been pointed out that the universal low energy long
distance physics of this Kondo model arises also in a spin chain
when a magnetic impurity is coupled to the end of a gapless
Heisenberg anti-ferromagnetic $J_1-J_2$ spin 1/2 chain, where
$J_1$ ($J_2$) is the (next) nearest neighbor coupling. When $J_2$
exceeds a critical value, the spin chain enters a gapped dimerized
regime and its relation to the Kondo model breaks down. Namely,
for $0\leq J_2 \leq J_2^c=0.2412$, the spin system is gapless and
it supports a Kondo regime
\cite{Laflorencie-Affleck,Sorensen-Affleck}. For $J_2>J_2^c$, the
system enters the gapped \emph{dimer regime}, where the ground
state takes a dimerised form; at the Majumdar-Ghosh \cite{MG}
point ($J_2=0.5$), the ground state becomes just a tensor product
of singlets. For $J_2>0.5$, incommensurability effects
\cite{pertinentAffleck} emerge.

Our aim in this paper is to use a true measure of entanglement to
fully characterize the unique features of the gapless Kondo regime in
the spin chain Kondo model. Namely, for this spin chain in the
Kondo regime: (i) we demonstrate that the impurity spin is indeed
maximally entangled with the Kondo cloud; (ii) we determine the
spatial extent of the Kondo screening length $\xi$ using only an
entanglement measure; (iii) we motivate an ansatz for its ground
state in the Kondo regime; (iv) we evidence the scaling of a true measure of
entanglement as pertinent parameters are varied. In order to
accomplish these tasks we device a DMRG approach enabling to
investigate the entanglement between a single spin and a pertinent
block of the chain, which may be applied in other contexts.
Finally, to better evidence the unique properties of the
entanglement in the Kondo regime we carry a parallel analysis of
the entanglement properties of this model in the gapped dimerised
regime. Using a true measure of entanglement to determine $\xi$
enables to exploit the peculiarities of the Kondo regime of a spin
chain to generate long range distance independent entanglement
usable for quantum communication tasks \cite{Sodano}.

A true measure of entanglement should satisfy a set of postulates
- for example, it should be non-increasing under local actions:
such a genuine measure does exist for two sub-systems of arbitrary
size even when their combined state is mixed, as it happens in
Kondo systems. This measure is the {\em negativity}
\cite{negativity} and it has been successfully used to quantify
the entanglement in a harmonic chain \cite{audenaert,kim-vedral}
and between distant regions of critical systems
\cite{hannu,alex-retzker}. For bipartite systems, negativity is
defined as $E=\sum_i|a_i|-1$, where $a_i$ denote the eigenvalues
of the partial transpose of the whole density matrix of the system
with respect to one of the two subsets of the given partition and
$|...|$ is the absolute value \cite{negativity}.

The paper is organized as follows: In section (\ref{lenght}) we define an entanglement healing
length for the spin chain Kondo model; section (\ref{methodology})
explains the DMRG-based approach we devised in order to compute its entanglement properties. In
section (\ref{results}) we show the remarkable scaling of a true measure of entanglement (i.e.negativity) in the
Kondo regime attainable by the Kondo spin chain when $0\leq J_2 \leq J_2^c=0.2412$; in addition, we motivate an ansatz
for the ground state of this chain in the Kondo regime. Section (\ref{conclusion}) is devoted to a summary of our
results and to a few concluding remarks.

\begin{figure}
\centering
    \includegraphics[width=7cm,height=6cm,angle=0]{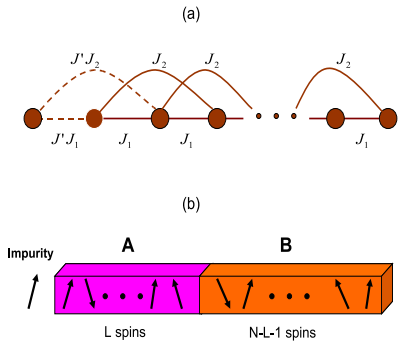}
    \caption{(Color online) a) Kondo Spin chain with next nearest neighbor Heisenberg interaction with one impurity at one end.
    b) The chain is divided into three parts, an impurity, a block $A$ and a block $B$. Entanglement is computed
    between the impurity and block $B$.}
     \label{fig1}
\end{figure}

\section{Measuring the Entanglement Healing Length} \label{lenght}
  The spin chain Kondo model \cite{kondo-Affleck} is defined by the Hamiltonian
\begin{equation}\label{Hamiltonian}
    H=J'(\sigma_1.\sigma_2+J_2\sigma_1.\sigma_3)+\sum_{i=2}^{N-1}\sigma_i.\sigma_{i+1}+J_2\sum_{i=2}^{N-2}\sigma_i.\sigma_{i+2},
\end{equation}
where $\sigma_i=(\sigma_i^x,\sigma_i^y,\sigma_i^z)$ is a vector of
Pauli operators at site $i$, $N$ is the total length of the chain,
$J_2$ is the next nearest neighbor coupling and the nearest
neighbor coupling $J_1$ has been normalized to $1$. The impurity
spin, located at one end of the chain, is accounted for by weaker
couplings to the rest of the system; in the following, see Fig.
\ref{fig1}a, both couplings $J_1$ and $J_2$ are weakened by the
same factor $J'$, which quantifies then the impurity strength.

To study the entanglement of the ground state we divide- see Fig.
\ref{fig1}b - all the spins of the chain in three different
groups: the impurity spin, block $A$, which contains the $L$ spins
next to the impurity ($L=0,1,...,N-1$) and block $B$ formed by the
remaining $N-L-1$ spins. We use negativity to fully characterize
the entanglement between the impurity and block $B$ in both the
gapless Kondo and the gapped dimerised regimes. We determine the
size of the block $A$ when the entanglement between the impurity
and block $B$ is almost zero; by this procedure we measure an
Entanglement Healing Length (EHL) $L^*$, i.e. the length of the
block $A$ which is maximally entangled with the impurity. We show
that, in the gapless Kondo regime, EHL scales with the strength of
the impurity coupling just as the Kondo screening length, $\xi$,
does. Thus, in the gapless regime of the Kondo spin chain, our
approach yields a method to detect the Kondo screening length
\cite{Sorensen-Affleck,Frahm,dots} based on a true measure of
entanglement. In addition, we {\em find} that entanglement, as
quantified by negativity, is a homogeneous function of two ratios:
$N/L^*$ and $L/N$, where $L$ is the size of the block $A$, i.e.
the block adjacent to the impurity, and $N$ is the length of the
whole chain. As a result, the entanglement in the Kondo regime is
essentially unchanged if one rescales all the length scales with
the EHL $L^*$.

\begin{figure}
\centering
    \includegraphics[width=7cm,height=6cm,angle=0]{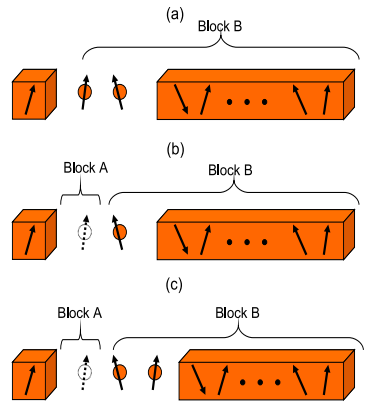}
    \caption{(Color online) a) DMRG representation of the state of the chain keeps two intermediate spins
    in ordinary computational basis and the left and right
    blocks in a truncated DMRG basis. b) The intermediate spin next to the impurity is traced out from the density matrix of the
    chain. This tracing is equivalent to adding the traced out spin
    to block $A$. c) The basis of the right block of DMRG representation is
transformed so that a single spin in the left side of the
    right block is represented in the computational basis while the state of the new
right block is given in a DMRG basis.}
     \label{fig2}
\end{figure}

\section{DMRG Analysis of Entanglement in the Spin Chain Kondo Model} \label{methodology}
  We use the Density Matrix Renormalization Group
(DMRG) \cite{white-DMRG} approach to compute the ground state of
the spin chain Kondo model. We analyze large chains, $N=250$, to
avoid finite size effects and take $N$ to be even to avoid
problems arising from accidental degeneracies.
In a DMRG approach the ground state of the system is partitioned
in terms of states of a left block, a right block
(not to be confused with blocks $A$ and $B$) and two intermediate
spins as shown in Fig. \ref{fig2}a. The states of the intermediate
spins are given in the computational
($|\uparrow\ra,|\downarrow\ra$) basis, while the states of the both
blocks are usually in some non-trivial truncated DMRG basis.
In this approach one has several representations for the ground state which
vary due to the number of spins in the left (right) block and it is possible
to go from one representation to the other by applying pertinent operators on each block. The main issue in the
DMRG is that the dimension of the left (right) block is kept constant independent of
whatever spins are there in that block. To have a fixed dimension for the left (right) block
we truncate the Hilbert space such that the amount of entanglement between the two parts
of the chain remains almost unchanged \cite{white-DMRG}. To have a precise results we need to sweep
all representations of the ground state for several times to get the proper basis for the left and the right blocks
of all representations. After some sweeps, when the ground state energy converges (we keep
states for which the error of the energy would be less than
$10^{-6}$), we pause to compute the entanglement. We take a
representation of the ground state in which the left block
contains just the single impurity spin and the right block
contains $N-3$ spins: as a result, the single impurity spin is
given in the computational basis and this allows us to compute the
negativity later.
 From this DMRG state, we trace
out the spins belonging to block $A$
 before computing the entanglement between the
impurity and block $B$ since it is most convenient to compute the
entanglement between the impurity and the block $B$: due to the
entanglement monogamy, this provides an equivalent information
about the entanglement of the impurity with the block $A$. Our
tracing procedure starts with the density matrix of the ground
state of the system in the representation shown in Fig.
\ref{fig2}a; at this stage, the number of spins in the block $A$
is zero (no spin has been traced out), all spins except the
impurity belong to the block $B$, and the entanglement between the
impurity and the block $B$ is maximal (i.e. $E=1$). Then, we trace
out the intermediate spin next to the impurity as shown in Fig.
\ref{fig2}b; this amounts to putting that spin into block $A$.
Finally, as shown in Fig. \ref{fig2}c, we transform the DMRG basis
of the right block so as to put a single spin at the left of the
right block in the computational basis, while the state of the new
right block is given in a DMRG basis. As a consequence, the
resulting density matrix  has the exact form of Fig. \ref{fig2}a
and we can continue the procedure to trace one spin at each step
(i.e., put more spins in the block $A$) and compute the
entanglement between impurity and block $B$.

\begin{figure}
\centering
    \includegraphics[width=8cm,height=7cm,angle=0]{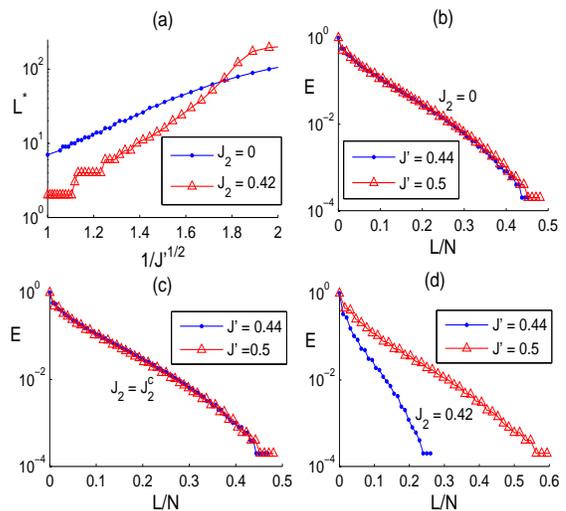}
    \caption{(Color online) a) $L^*$ vs. $1/\sqrt{J'}$ for both Kondo ($J_2=0$) and dimer regime ($J_2=0.42$).
    b) Entanglement vs. $L/N$ for fixed $N/L^*=4$ when $J_2=0$.
    c) Entanglement vs. $L/N$ for fixed $N/L^*=4$ at the critical point $J_2=J_2^c$.
    d) Entanglement vs. $L/N$ for fixed $N/L^*=4$ in the dimer regime ($J_2=0.42$).}
     \label{fig3}
\end{figure}

\section{Scaling of Negativity and Ansatz for the Ground State in the Kondo Regime} \label{results}
We find that there is an EHL $L^*$ so that, for $L>L^*$, the
entanglement between the impurity and block $B$ is almost zero:
$L^*$ provides us with an estimate of the distance for which the
impurity is mostly entangled with the spins contained in block
$A$. For large chains ($N>200$) in the Kondo regime, one finds
that $L^*$ is almost independent of $N$ and depends only on $J'$.
In the Kondo regime, i.e. for $J_2<J_2^c$, $L^*$ depends on $J'$
just as the Kondo screening length $\xi$ does
\cite{Laflorencie-Affleck,Sorensen-Affleck}; for small $J'$,
$L^*\propto e^{\alpha/\sqrt{J'}}$, where $\alpha$ is a constant.
We plot $L^*$ as a function of $1/\sqrt{J'}$ in Fig. \ref{fig3}a.
In a semilogarithmic scale, the straight line plot exhibited in
the Kondo regime ($J_2=0$) shows that $L^*$ may be regarded as the
Kondo screening length. Moreover, the nonlinearity of the same
plot in the dimer regime ($J_2=0.42$), especially for small $J'$,
shows that, here, no exponential dependence on $1/\sqrt{J'}$
holds.

 We observe also a remarkable scaling of negativity in the Kondo regime.
This scaling may be regarded as yet another independent evidence
of the fact that $L^*$ is indeed the Kondo length $\xi$. In
general, the entanglement $E$ between the impurity and block $B$
is a function of the three independent variables, $J',L$ and $N$
which, due to the one to one correspondence between $J'$ and
$L^*$, can be written as $E(L^*,L,N)$. We find that, in the Kondo
regime, $E=E(N/L^*,L/N)$.  To illustrate this, we fix the ratio
$N/L^*$ and plot the entanglement in terms of $L/N$ for different
values of $J'$ (or equivalently $L^*$) for $J_2=0$ (Fig.
\ref{fig3}b) and for $J_2=J_2^c$ (Fig. \ref{fig3}c). The complete
coincidence of the two plots in Figs. \ref{fig3}b and c shows
that, in the Kondo regime, the spin chain can be scaled in size
without essentially affecting the entanglement as long as $L^*$ is
also scaled. In the dimer regime the entanglement stays a function
of three independent variables, i.e. $E=E(L^*,L,N)$, and, as shown
in Fig. \ref{fig3}d, the entanglement does not scale with $L^*$.
In our approach, the Entanglement Healing Length $L^*$ may be
evaluated in both the Kondo and the dimer regime: the scaling
behavior, as well as the dependence of $L^*$ on $J'$,
discriminates then between the very different entanglement
properties exhibited by the spin chain Kondo model as $J_2$
crosses $J_2^c$.

We defined $L^*$ such that there is no entanglement between the
impurity and block $B$ when block $A$ is made of $L^*$ spins.
Conventional wisdom based on previous renormalization group
analysis suggests that, in both regimes, the impurity and the
block $A$ of length $L^*$ form a pure entangled state, while block
$B$ is also in a pure state. This is indeed approximately true in
the dimer regime (exactly true for $J_2=0.5$) but it turns out to
be dramatically different in the Kondo regime. To check this, we
computed the von Neumann entropy of the block $B$ when block $A$
has $L^*$ spins and found it to be non zero. Thus, the blocks $A$
and $B$ are necessarily entangled in the Kondo regime as there is
no entanglement between the impurity and $B$.
 In fact, after a distance
 $L^*$,
the impurity is "screened" i.e, the block $B$ feels as if it is
part of a conventional gapless chain and has a diverging von
Neumann entropy. The Kondo cloud is maximally entangled with the
impurity as well as being significantly entangled with block $B$.
Based on the above, a simple ansatz for the ground state $|GS\ra$
in the Kondo regime is provided by
\begin{equation}\label{state_kondo}
    |GS\ra=\sum_i \alpha_i
    \frac{|\uparrow\ra|L_i^\uparrow(J')\ra-|\downarrow\ra|L^\downarrow_i(J')\ra}{\sqrt{2}}\otimes
    |R_i(J')\ra,
\end{equation}
where $\alpha_i$ are constants,
$\{|L_i^\uparrow(J')\ra,|L^\downarrow_i(J')\ra\}$ and
$\{|R_i(J')\ra\}$ are sets of orthogonal states on the cloud and
the remaining system, respectively. At the fixed point
$J'\rightarrow 0$ all spins except the impurity are included in
$|L_i^\uparrow(J')\ra$ and $|L_i^\downarrow(J')\ra$. At
$J'\rightarrow 1$, very few spins are contained in
$|L_i^\uparrow(J')\ra$ and $|L_i^\downarrow(J')\ra$ while
$\{|R_i(J')\ra\}$ represents most of the chain.

\begin{figure}
\centering
    \includegraphics[width=8cm,height=7cm,angle=0]{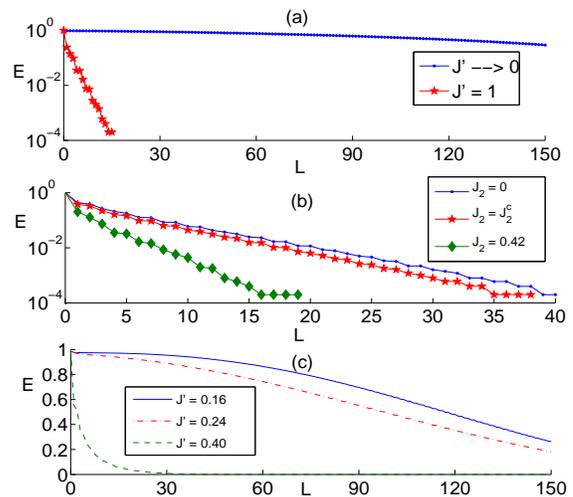}
    \caption{(Color online) a) Entanglement vs. $L$ for the two fixed points $J'\rightarrow 0$
     and $J'\rightarrow 1$ in the Kondo regime $J_2=0$.
    b) Exponential decay of entanglement in terms of $L$ in a chain of length $N=250$ for $J'=0.6$.
    c) Non-exponential decay for small $J'$ in the dimer regime $J_2=0.42$.}
     \label{fig4}
\end{figure}

  For what concerns the mere evaluation of the amount of entanglement as $J'$ is varied, we first plot, in Fig.
\ref{fig4}a, the negativity as a function of $L$ near the two
fixed points, i.e. $J'\rightarrow0$ and $J'\rightarrow1$,
accessible in the Kondo regime: as expected, near $J'\rightarrow0$
(i.e, for large values of the Kondo screening length), the
entanglement remains large for rather large values of $L$ while,
for $J'\rightarrow 1$ (i.e. for a very small cloud) it decreases
rapidly with $L$. Fig. \ref{fig4}a (semilogarithmic) shows that,
also at the extreme limits $J'\rightarrow0$ and $J'\rightarrow1$,
the entanglement decays exponentially with $L$ since this a
characteristic mark of the entanglement in the Kondo regime. This
exponential decay of entanglement is absent in the dimer regime:
Fig. \ref{fig4}b shows that, here, only for rather large $J'$, the
entanglement decays exponentially with $L$ while, for small $J'$,
the entanglement between the impurity and block $B$ decays {\em
slower than an exponential} as a function of $L$ exhibiting even a
plateau at short distances. This latter feature is evidenced in
Fig. \ref{fig4}c, and is consistent with the emergence, for small
$J'$, of long range valence bonds between the impurity and far
spins as a consequence of the onset of the dimerised ground state
\cite{Sorensen-Affleck}. In fact, when $J'$ is small, $J_2J'$
becomes much less than $J_2^c$ and the impurity forms valence
bonds with distant spins, while the other spins, since for them
$J_2>J_2^c$, form a valence bond with their nearest neighbor to
preserve the dimerised nature of the ground state: as a result,
the impurity shares less entanglement with nearby spins and
fulfills its capacity of entanglement forming valence bonds with
the more distant spins in the chain.

\section{Summary and Concluding Remarks} \label{conclusion}

  To summarize, we analyzed the Kondo spin
chain model from the viewpoint of a genuine entanglement measure,
namely the negativity. This readily shows that the impurity spin
is indeed maximally entangled with the Kondo cloud. We used
negativity to provide an {\em independent} method to determine the
Kondo screening length and to provide a characterization of the
ground state of the Kondo spin chain in the Kondo regime. We found
that, not only is the Kondo regime of this model distinct from the
gapless phases probed to date using the von Neumann entropy, but
the form of the entanglement -- a spin and a block in a mixed
state -- is also distinctive. We deviced a DMRG approach enabling
to compute the entanglement between the impurity and a block of
spins located at the other side of the chain for different lengths
of the block. We showed that, in the Kondo regime, the EHL $L^*$
scales with the impurity coupling $J'$ just as the Kondo length
does: in other words, the impurity, though not entangled with any
individual spin, is shown to be entangled with the totality of the
spins within the Kondo cloud- whose size is measured by $L^*$- and
disentangled from the rest. Our measure of the entanglement in the
Kondo regime led us to formulate an ansatz for the ground state of
the Kondo spin chain for $J_2<J_2^c$. Our approach also shows
that, in the Kondo regime, the entanglement scales exponentially
with $L/L^*$ and that, in the gapped dimer regime, though it is
still possible to define an EHL, the impurity-block entanglement
is usually smaller and has no characteristic length scale.

  We thank H. Wichterich and V. Giovannetti for valuable discussions.
AB and SB are supported by the EPSRC, QIP IRC (GR/S82176/01), the
Royal Society and the Wolfson Foundation. PS was partly supported
by the ESF Network INSTANS.

\end{document}